\documentstyle[preprint,floats,aps,psfig]{revtex}

\tighten

\begin{document}

\draft

\title{Adjoint ``quarks'' on coarse anisotropic lattices: \\
Implications for string breaking in full QCD}

\author{Karn Kallio and Howard D. Trottier}
\address{Physics Department, Simon Fraser
University, Burnaby, B.C., Canada V5A 1S6}

\maketitle

\begin{abstract}
A detailed study is made of four dimensional SU(2) gauge theory with static
adjoint ``quarks'' in the context of string breaking.
A tadpole-improved action is used to do simulations on lattices with coarse
spatial spacings $a_s$, allowing the static potential to be probed at large
separations at a dramatically reduced computational cost. Highly anisotropic
lattices are used, with fine temporal spacings $a_t$, in order to assess
the behavior of the time-dependent effective potentials.
The lattice spacings are determined from the potentials for quarks in the
fundamental representation. Simulations of the Wilson loop in the adjoint
representation are done, and the energies of magnetic and electric
``gluelumps'' (adjoint quark-gluon bound states) are calculated, which set
the energy scale for string breaking. Correlators of gauge-fixed static
quark propagators, without a connecting string of spatial links, are analyzed.
Correlation functions of gluelump pairs are also considered; similar
correlators have recently been proposed for observing string breaking in
full QCD and other models. A thorough discussion of the relevance of Wilson
loops over other operators for studies of string breaking is presented,
using the simulation results presented here to support a number of
new arguments.
\end{abstract}

\section{Introduction and Motivation}

Simulations of lattice QCD are increasingly dedicated to the goal of
including the effects of sea quarks on many observables.
One of the most distinctive signatures of sea quarks should be the
elimination of the confining potential between widely separated valence
quarks. Quenched simulations have demonstrated that
color-electric field lines connecting a static quark and antiquark
are squeezed into a narrow tube or ``string'' \cite{MichaelTube}.
In full QCD however the flux-tube should be unstable against fission
at large separations $R$, where sea quarks should materialize from
the vacuum and bind to the heavy quarks to form a pair of
color-neutral bound states.

It is perhaps surprising that some controversy persists in the literature
as to whether ``string breaking'' has actually been observed in lattice
simulations, and what techniques are required in order to convincingly
demonstrate this phenomenon. This is despite extensive large scale
simulations in unquenched QCD by several collaborations
\cite{Glassner,Aoki,Talevi,SESAMTchiL,Allton}.
This problem has come under renewed attack in the last
few years with several new viewpoints emerging as to the
underlying cause of the difficulty in observing string breaking, and
suggestions as to the optimal approach for resolving this problem
\cite{Gusken,Burger,DeTar,Drummond,PhilScalar,Knechtli,%
HDT,Gliozzi,StringReviews}. These ideas have stimulated new work
on string breaking in simulations of full QCD
\cite{Lacock,Eichten,MichaelNew},
and on a number of models that may shed light on string breaking
\cite{Koniuk,Stephenson,PhilAdjoint3D,PhilAdjoint4D,%
Deldar,Leinweber}.

One suggestion to arise in the literature is that the Wilson
loop operator, which has typically been used to study the static potential
between heavy quarks, has a very small overlap with the true
ground state of the system at large $R$, and hence is not suited
to studies of string breaking \cite{Aoki,Gusken,Drummond}.
This has led to the consideration of other operators to
study string breaking between static quarks, especially operators that
explicitly generate light valence quarks at the positions of
the heavy quarks \cite{PhilScalar,Knechtli}. We note that another
distinguishing feature of the operators proposed in Refs.
\cite{PhilScalar,Knechtli} is that they generate correlation functions
which receive disconnected contributions, where a pair of heavy-light
bound states propagate independently of one another.

Another point of view was raised by one of us in Ref. \cite{HDT},
where it was suggested that string breaking can indeed be seen
using Wilson loops, but that it is essential to propagate the
trial states over Euclidean times $T$ of about 1~fm, the
characteristic scale associated with hadronic binding.
In contrast, typical studies of the static potential in
unquenched QCD have been done on lattices with relatively ``fine''
spacings, limiting the propagation times to well under 1~fm,
due to the high computational overhead. The use of coarse lattices
enables much more efficient simulations of the static potential
at the large scales relevant to string breaking.
This was demonstrated in Ref. \cite{HDT} where an improved action
was used to observe string breaking on coarse lattices, using
Wilson loops in unquenched QCD in three dimensions,
at dramatically reduced computational cost.

In this paper we consider a number of the issues raised in the recent
literature on string breaking, by doing simulations of quenched
lattice QCD with valence ``quarks'' in the adjoint representation
of the color group. There is a long history of lattice simulations
of QCD with adjoint matter fields, which exhibits much of the
physics of confinement of real QCD, and which also has a connection
to supersymmetric physics \cite{MichaelAdj}. In particular an analogue of
string breaking should occur in this model. For instance, the confining
flux-tube between a pair of heavy adjoint quarks
should be unstable against fission at large $R$, where gluons can
materialize from the sea to bind to the heavy quarks, forming a pair of
color-neutral bound states dubbed ``gluelumps'' \cite{MichaelAdj}.
Hence the potential $V_{\rm adj}(R)$ between a pair of static adjoint
quarks in quenched QCD should approach a constant at large $R$
\begin{equation}
   V_{\rm adj}(R\to\infty) = 2 M_{Qg} ,
\label{Vbreak}
\end{equation}
where $M_{Qg}$ is the energy of the lightest gluelump.
Once again, despite much effort
\cite{MichaelAdj,Mawhineey,Woloshyn,HDTAdjoint,Poulis,Bali,Deldar},
string breaking using Wilson loops has not been seen in this model.
However it has been suggested very recently that string breaking
for adjoint quarks can be seen using correlators that explicitly
generate valence gluons \cite{Stephenson,PhilAdjoint3D,PhilAdjoint4D}.

Here we undertake a detailed study of four dimensional SU(2) gauge theory
with static adjoint quarks (this work was reported in unpublished form in
Ref. \cite{KarnThesis}). We use a tadpole-improved gluon action to do
simulations on lattices with coarse spatial spacings $a_s$. We use highly
anisotropic lattices, with fine ``temporal'' spacings $a_t$, in order to
make a careful study of the time-dependent effective potentials.
One lattice used here has $a_s = 0.36$~fm and $a_t=0.10$~fm, which provides
an increase in computational efficiency of some two orders of magnitude
compared to simulations of adjoint quarks that were done in
Ref. \cite{PhilAdjoint4D} using an unimproved action on lattices
with spacings of about 0.1~fm.

The lattice spacings are first determined from the potentials for quarks in the
fundamental representation. We then study the Wilson loop in the adjoint
representation, and the masses of magnetic and electric gluelumps, which
set the energy scale for string breaking according to Eq. (\ref{Vbreak}).
We consider correlators of gauge-fixed static quark propagators,
without the string of spatial links that is found in the Wilson
loop; this is similar to correlators proposed in Refs. \cite{Burger,DeTar}
as alternatives to the Wilson loop for observing string breaking.
We also consider correlation functions for a pair of gluelumps,
corresponding to states that explicitly contain valence gluons.

We find that adjoint quark string breaking is extremely difficult to
observe using Wilson loops, because of a very strong suppression of the
signal, due to approximate Casimir scaling of the static potential. Despite
the considerable computational advantage provided by using coarse lattices,
we are limited to propagation times well below 1~fm. It is clear from our
results that plateaus in the effective mass plots cannot be established at
such short times. However it is also clear than a progressive ``flattening''
of the adjoint potential occurs as the propagation time is increased.

By contrast saturation is readily observed in the effective potential
naively constructed from the gluelump-pair correlator. Similar results for
some of the observables studied here were recently obtained on ``fine''
lattices using unimproved actions
\cite{Stephenson,PhilAdjoint3D,PhilAdjoint4D}. The adjoint quark
results in Refs. \cite{Stephenson,PhilAdjoint3D,PhilAdjoint4D}
were interpreted as providing support for the picture
\cite{Aoki,Gusken,Drummond} that Wilson loops are not suitable for studies
of string breaking, while suggesting that string breaking can readily be
observed using operators that explicitly generate states containing
light valence particles \cite{PhilScalar,Knechtli}.

However we suggest instead that one must be more careful in defining
the goal of ``observing string breaking'' in Euclidean time correlation
functions. In particular we note that operators which
explicitly generate light valence quarks, such as proposed in
Refs. \cite{PhilScalar,Knechtli,Stephenson,PhilAdjoint3D,PhilAdjoint4D},
will {\it automatically\/} exhibit potentials that saturate, even in
{\it quenched\/} QCD (here considering the theory with fundamental
representation quarks). This is especially clear when the correlator has
disconnected contributions. This can also be seen from results presented
in Ref. \cite{Koniuk}, where correlation functions of exactly this form
were studied in quenched QCD with fundamental representation quarks.
If one goal of string breaking studies is to observe a distinctive feature
of the effects of sea quarks, then clearly one must use an observable
that distinguishes between the quenched and unquenched theories.

We further argue that using an operator such as the Wilson loop (which
might be thought of as creating ``initial'' states containing only heavy
valence quarks) is well motivated by making an analogy with the process
of hadronization, which is the physical process that is most often alluded
to in the context of ``string breaking.'' In hadronization the initial state
consists of just two valence quarks at small separations, which then separate
in real time, leading to the creation of light valence quarks at separations
around 1--2~fm. The small overlap of the Wilson loop  with the broken string
state at much larger separations may be an irreducible feature of using
Euclidean time correlators to mimic hadronization; a nonlocal state of two
widely separated heavy quarks connected by a string is bound to have a small
overlap with the state consisting of two widely separated heavy-light mesons.
On the other hand, we find that the overlap of the Wilson loop with the broken
string state is appreciable in a range of separations around the point at
which string breaking actually occurs. Simulations of Wilson loops at very
large separations, where the overlap becomes very small, are not relevant
to hadronization, since in this physical process the original quarks never
get to such points with the string intact.

If one considers the static potential for quark separations around 1--2~fm,
which are physically motivated by the analogy with hadronization, then
``string breaking'' does indeed appear to be accessible in full QCD using
Wilson loops. The important observation here is that propagation times of
about 1~fm must be attained in order to resolve the broken string
state \cite{HDT}. This can be more readily achieved by using improved actions
to do simulations on coarse lattices; we estimate that an increase in
computational efficiency of some two orders of magnitude over most recent
studies in full QCD can reasonably be achieved. This viewpoint is also
supported by the results presented here, given the behavior
of the time dependent effective masses, even though actual string breaking
in adjoint Wilson loops is not resolved. In this connection, we draw attention
below to the somewhat surprising fact that string breaking in quenched QCD
with adjoint quarks actually has a {\it much\/} higher computational burden
than in unquenched QCD with real quarks. In our estimation all of the evidence
supports the view that string breaking is accessible in large scale
simulations of the Wilson loop in full QCD.

The rest of this paper is organized as follows. In Sect.~II we
present the details of the improved gluon action, and the
construction of the various correlation functions to be studied.
The results of the simulations are presented in Sect.~III,
where each correlator is considered in turn. Finally, we
present some further discussion and conclusions in Sect.~IV.

\section{Action and Observables}

The tadpole-improved SU(2) action on anisotropic lattices
used here was previously studied in Ref. \cite{NormSU2}, following
earlier work in SU(3) color \cite{LepAn,ColinAn}
\begin{eqnarray}
   {\cal S} &=& - \beta \sum_{x,\,s>s^\prime}
   \xi_0 \,\left\{
   \frac{5}{3}  \frac{P_{ss^\prime}}{u_s^4}
   - \frac{1}{12} \frac{R_{ss^\prime}}{u_s^6}
   - \frac{1}{12} \frac{R_{s^\prime s}}{u_s^6} \right\}
 \nonumber \\
   && - \beta \sum_{x,\,s}
   {1 \over \xi_0} \,\left\{
   \frac{4}{3}  \frac{P_{st}}{u_s^2 u_t^2}
   - \frac{1}{12} \frac{R_{st}}{u_s^4 u_t^2} \right\} ,
\label{Simp}
\end{eqnarray}
where $P_{\mu\nu}$ is one half the trace of the $1\times1$ Wilson
loop in the $\mu\times\nu$~plane, $R_{\mu\nu}$ is one half the
trace of the $2\times1$ rectangle in the $\mu\times\nu$~plane,
and where $\xi_0$ is the bare lattice anisotropy,
\begin{equation}
   \xi_0 = \left( {a_t \over a_s} \right)_{\rm bare} .
\end{equation}
This action has rectangles $R_{ss^\prime}$ and $R_{st}$ that extend at most
one lattice spacing in the time direction. This has the advantage of
eliminating a negative residue high energy pole in the gluon propagator
that would be present if $R_{ts}$ rectangles were included.
``Diagonal'' correlation functions computed from this action thus
decrease monotonically with time, which is very important for
our purposes. The leading discretization errors in this action are thus
of $O(a_t^2)$ and $O(\alpha_s a_s^2)$.

On an anisotropic lattice one has two mean fields $u_t$ and
$u_s$. Here we define the mean fields using the measured
values of the average plaquettes. Since the lattice
spacings $a_t$ in our simulations are small, we adopt the following
prescription \cite{NormSU2,LepAn,ColinAn} for the mean fields
\begin{equation}
   u_t \equiv  1 ,
   \quad
   u_s = \langle P_{s s'} \rangle^{1/4} .
\end{equation}

Observables in various representations of the gauge group can be
easily computed from the measured values of the fundamental
representation link variables using relations amongst the group
characters. The Wilson loop $W_j$ in the $j$-th
representation is defined by
\begin{equation}
   W_j \equiv {1\over 2j+1} \mbox{Tr}
       \left\{ \prod_{l \in L} {\cal D}_j[U_l] \right\} ,
\end{equation}
where ${\cal D}_j[U_l]$ is the $j$-representation of the link
$U_l$, and $L$ denotes the path of links in the Wilson loop.
In the case of the adjoint Wilson loop of interest here, we have
\begin{equation}
   W_1(T,R) = \case{1}{3} \left( 4 \, \vert W_{1/2}(T,R) \vert^2 - 1 \right) ,
\end{equation}
as can also be seen by using an explicit form for the
adjoint representation matrices \cite{MichaelAdj}
\begin{equation}
   {\cal D}_1^{ab}[U] = \case12 \mbox{Tr}( \sigma^a U \sigma^b U^\dagger ) ,
\label{D1}
\end{equation}
and making use of the identity
$\sigma^a_{ij} \sigma^a_{kl} = 2(\delta_{il} \delta_{jk}
-\frac12 \delta_{ij} \delta_{kl})$.

To enhance the signal-to-noise we make an analytical integration
on time-like links \cite{VarRed}
\begin{equation}
   \int d[U_l] {\cal D}_j[U_l] e^{-\beta S} =
   {I_{2j+1}(\beta k_l) \over I_1(\beta k_l)} {\cal D}_j[V_l]
   \int d[U_l] e^{-\beta S}
\label{varred}
\end{equation}
where $k_l V_l$ denotes the sum of the $1\times 1$ staples and $2 \times 1$
rectangles connected to the time-like link $U_l$ [det$(V_l) \equiv 1$].
This variance reduction was applied to time-like links for the Wilson
loops and gluelump correlators. Equation (\ref{varred}) assumes that
a given link appears linearly in the observable, hence we can only apply it
to Wilson loops with $R>2$, because of the rectangles $R_{st}$ that appear
in the improved action.

An iterative fuzzing procedure \cite{APE} was used to increase
the overlap of the Wilson loop and gluelump operators with
the lowest-lying states. Fuzzy link variables $U^{(n)}_i(x)$ at the
$n$th step of the iteration were obtained from a linear combination of
the link and surrounding staples from the previous step
\begin{equation}
   U^{(n)}_i(x) = U^{(n-1)}_i(x)
   + \epsilon \sum_{j \neq \pm i}
   U^{(n-1)}_j(x) U^{(n-1)}_i(x+\hat\jmath)
   U_j{^{(n-1)}}^\dagger(x+\hat\imath)  ,
\label{fuzz}
\end{equation}
where $i$ and $j$ are purely spatial indices, and where the
links were normalized to $U^\dagger U = I$ after each iteration.
Operators were constructed by using the fuzzy spatial link variables in
place of the original links. Typically the number of
iterations $n$ and the parameter $\epsilon$ were chosen around
$(n,\epsilon)=(10,0.04)$ for Wilson loops and
$(n,\epsilon)=(4,0.1)$ for gluelumps.

The gauge-invariant propagator $G(T)$ of a gluelump can be constructed
by coupling a static quark propagator $Q_T$ (product of temporal links)
of time-extent $T$ to spatial plaquettes $U_0$ and $U_T$
located at the temporal ends of the line \cite{MichaelAdj}
\begin{equation}
   G(T) = \mbox{Tr} \left( U_0 \sigma^b \right)
         {\cal D}_1^{ab} \left[ Q_T \right]
          \mbox{Tr} \left( U_T^\dagger \sigma^b \right) .
\label{Gcorr}
\end{equation}
Both magnetic and electric gluelump propagators were analyzed, by
choosing appropriate linear combinations of spatial plaquettes at the
ends of the static propagator \cite{MichaelAdj}. For the magnetic gluelump
a sum of four plaquettes in a particular spatial plane is used, the sum
being invariant under lattice rotations about an axis perpendicular the
plane. For the electric gluelump a sum of eight plaquettes lying
in two planes is used, the sum being invariant under rotations about an
axis that lies in both planes. $G(T)$ can be expressed in terms of
fundamental representation link variables using Eq. (\ref{D1})
\begin{equation}
   G(T) = 2 \mbox{Tr} \left( U_0 Q_T U_T^\dagger Q_T^\dagger \right)
        - \mbox{Tr} \left( U_0 \right) \mbox{Tr} \left( U_T \right) .
\end{equation}

We also study the correlation between two gluelump propagators, measuring
the expectation value of the operator
\begin{equation}
   G_{GG}(T,R) \equiv G^\dagger(T;R) G(T;0) ,
\label{GGPair}
\end{equation}
where static quark propagators $Q_{T;0}$ and $Q_{T;R}$ of time-extent
$T$, and separated by a spatial distance $R$, are used in
$G(T;0)$ and $G(T;R)$, respectively. We also compute the off-diagonal
entries in the gluelump pair-Wilson loop mixing matrix, given by the
expectation value of the operators
\begin{equation}
   G_{GW}(T,R) \equiv
   \mbox{Tr} \left( U_{0;0} \sigma^a \right)
   {\cal D}_1^{ab} \left[
   Q_{T;0} \, \Gamma_{T;R} \, Q_{T;R}^\dagger \right]
   \mbox{Tr} \left( U_{0;R} \sigma^b \right) ,
\label{GGW}
   \end{equation}
and
\begin{equation}
   G_{WG}(T,R) \equiv
   \mbox{Tr} \left( U_{T;R} \sigma^b \right)
   {\cal D}_1^{ab} \left[
   Q_{T;R}^\dagger \, \Gamma_{0;R}^\dagger \, Q_{T;0} \right]
   \mbox{Tr} \left( U_{T;0} \sigma^b \right) .
\label{GWG}
\end{equation}
$\Gamma_{0;R}$ and $\Gamma_{T;R}$ are products of (fuzzy) links connecting
the spatial sites of the heavy quarks at times zero and $T$ respectively.
The plaquettes $U_{0;0}$ and $U_{0;R}$ in the case of $G_{GW}$ for example
are connected to the ends of the static propagators at time zero.

Finally, as an alternative to the Wilson loop, we compute correlators of
gauge-fixed static quark propagators separated by a distance $R$,
given by expectation values of the operator
\begin{equation}
    G_{\rm Poly}(T,R) =
    \mbox{Tr}\left( {\cal D}_1[ Q_{T;0} ] \right)
    \mbox{Tr}\left( {\cal D}_1[ Q_{T;R} ] \right) .
\label{GPoly}
\end{equation}
This operator is similar to the Wilson loop in that it has only
heavy quark propagators. It has been suggested \cite{Burger,DeTar}
that this type of operator may have a larger overlap with the broken
string state, since it does not have an explicit string of spatial
links connecting the heavy quarks, unlike the Wilson loop.
We measured $G_{\rm Poly}$ in lattice Coulomb gauge, where
$\sum_{i=1}^3 [U_i(x)-U_i(x-\hat\imath)]=0$, which we implemented
using an iterative steepest ascent algorithm with fast Fourier
acceleration \cite{DaviesFFT}.

\section{Results}

\subsection{Lattice parameters and fundamental representation potentials}

Four lattices were studied in order to check the physical results
for dependence on lattice spacing and input anisotropy.
The four sets of simulation parameters are listed in
Table~\ref{table:params}. Roughly 40,000 measurements
were made for the observables on all lattices, skipping 10 configurations
between measurements (which results in very small autocorrelation times).

\begin{table}
\begin{center}
\begin{tabular}{cccccccc}
$\beta$ & $\xi_0$      & $u_s$   & Volume
        & $\xi_{\rm ren} / \xi_0$
        & $a_s$ (fm)   & $a_t$ (fm)
        & $\Delta V(\sqrt3 a_s)$  \\
\hline
0.848   & 0.276     & 0.7933     & $10^3 \times 20$
        & 1.02(1)   & 0.361(8)   & 0.102(2)  &  0.078(2)  \\
0.848   & 0.125     & 0.8432     & $8^3  \times 30$
        & 1.17(1)   & 0.494(20)  & 0.072(3)  &  0.154(2)  \\
0.600   & 0.125     & 0.7947     & $8^3  \times 30$
        & 1.14(2)   & 0.606(40)  & 0.086(6)  &  0.220(2)  \\
0.500   & 0.125     & 0.7648     & $8^3  \times 30$
        & 1.12(1)   & 0.689(30)  & 0.096(4)  &  0.231(2)  \\
\end{tabular}
\end{center}
\caption{Simulation parameters for the four lattices, and
measured values of some lattice quantities. The bare anisotropies $\xi_0$
and the mean fields $u_s$ for tadpole improvement are shown (where
$u_t \equiv 1$), along with the lattice volume in each case.
Measured values of the lattice anisotropies $\xi_{\rm ren}$
are compared to the input anisotropies, as discussed in the text.
Simulation results for the spatial and temporal spacings
$a_s$ and $a_t$ are given, as well as the relative errors
$\Delta V$ in the off-axis potentials at $R = \sqrt3 a_s$.}
\label{table:params}
\end{table}

We note that lattices with coarse spatial spacings $a_s$
were deliberately chosen, in an effort to probe the potentials
at the longest physical quark separations possible, for
the least computational cost. Lattice spacings around
0.4~fm have been studied by a number of authors
(see e.g. Refs. \cite{GPL,ColinAn,NormSU2}).
Using the lattice with $a_s=0.36$~fm here provides an increase in
computational efficiency of some two orders of magnitude
compared to the simulations of adjoint quarks done on lattices
with ``fine'' spacings in Ref. \cite{PhilAdjoint4D}.
Still coarser spacings were also considered here and,
although one would not necessarily advocate the use of such
lattices for precision calculations, it is worthwhile to employ
them in an effort to glean some information about string breaking,
where much remains to be understood.

We first present results for the fundamental representation Wilson loop,
which are used to measure the renormalized lattice anisotropy
$\xi_{\rm ren}$ and to set the lattice spacing $a_s$.
These results demonstrate fairly good scaling and rotational symmetry
restoration, as well as fairly small renormalizations of the input
anisotropy, thanks to the tadpole improvement of the action.

The renormalized anisotropy is determined by comparing
the static potential $a_t V_{xt}$, computed in units of $a_t$ from Wilson
loops $W_{xt}$ where the time axis is taken in the direction of
small lattice spacings, with the potential $a_s V_{xy}$ computed
from Wilson loops $W_{xy}$ with both axes taken in the direction of large
lattice spacings \cite{NormSU2,LepAn,ColinAn,Klassen}.
The anisotropy is determined after an unphysical
constant is removed from the potentials, by subtraction of
the simulation results at two different radii
\begin{equation}
  \xi_{\rm ren} =
  { a_t V_{xt}(R_2) - a_t V_{xt}(R_1)   \over
    a_s V_{xy}(R_2) - a_s V_{xy}(R_1)  } .
\label{xiren}
\end{equation}
The anisotropies determined with $R_1=\sqrt2a_s$ and $R_2=2a_s$
are shown in Table \ref{table:params}; results obtained with $R_1 = a_s$
are in excellent agreement with these estimates. The renormalization of
the anisotropy is fairly small in all cases, especially as compared to
the very large renormalizations for unimproved actions on
lattices with comparable spacings \cite{NormSU2}.

\begin{figure}
\vspace{0.25cm}
\centerline{\psfig{file=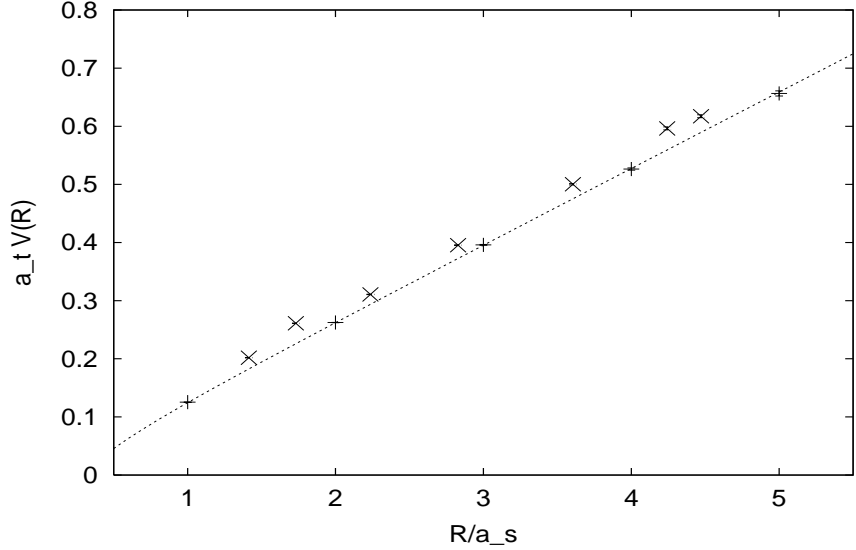,height=7.5cm,width=12.0cm}}
\vspace{0.50cm}
\caption{Fundamental representation potential for the action
with $a_s=0.49$~fm: on-axis points ($+$), off-axis points ($\times$).
The dotted line shows the results of a fit of the on-axis points
to Eq. (\ref{Vfit}).}
\label{fig:FWilFit}
\end{figure}

\begin{figure}
\vspace{0.25cm}
\centerline{\psfig{file=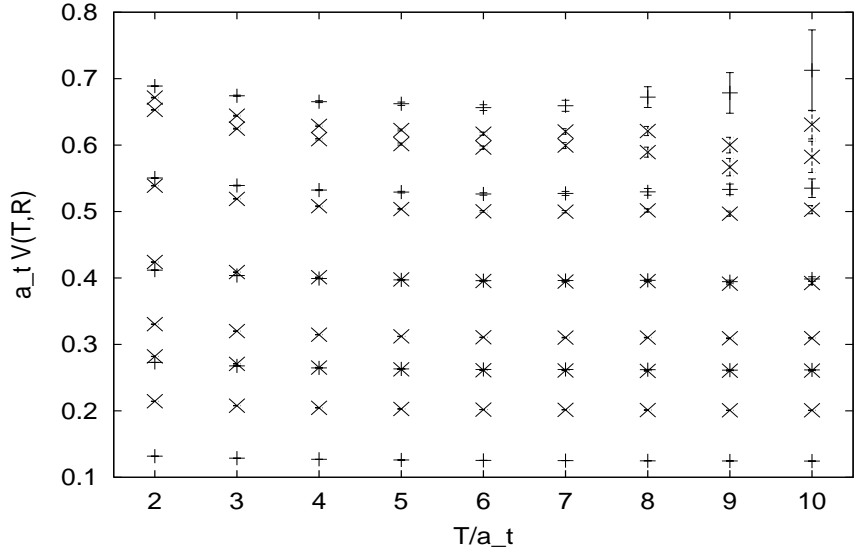,height=7.5cm,width=12.0cm}}
\vspace{0.50cm}
\caption{Time-dependent effective mass plot for the fundamental
potential from the lattice with $a_s=0.49$~fm and $a_t=0.072$~fm.
Each roughly horizontal line of points shows the effective mass at one
separation $R$: on-axis points ($+$), off-axis points ($\times$).}
\label{fig:FWilEmass}
\end{figure}

\begin{figure}
\vspace{0.25cm}
\centerline{\psfig{file=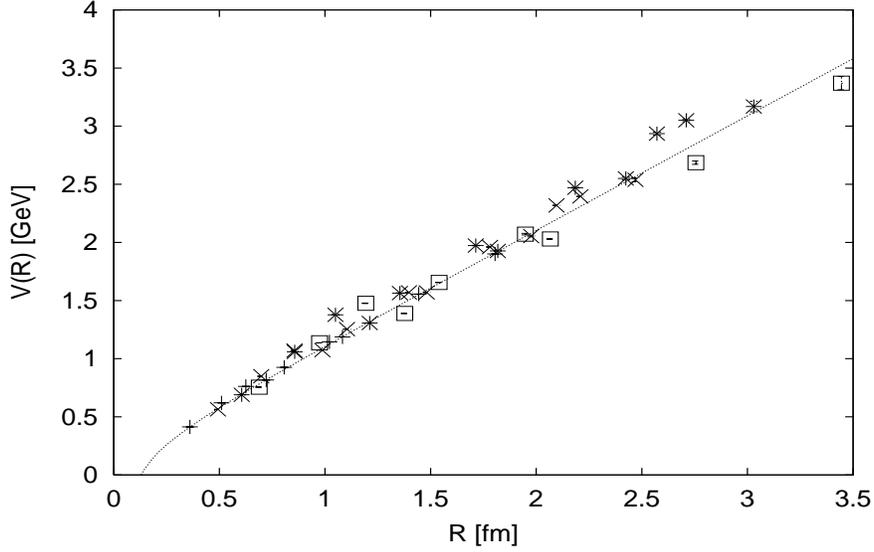,height=7.5cm,width=12.0cm}}
\vspace{0.50cm}
\caption{Fundamental representation potentials in physical units
from all four lattices:
$a_s=0.36$~fm ($+$), $a_s=0.49$~fm ($\times$),
$a_s=0.61$~fm ($*$), and $a_s=0.69$~fm ($\Box$).
An additive renormalization in the energies has been adjusted so
that the potentials agree at $R \approx 0.5$~fm.
The dotted line is the best fit to Eq.~(\ref{Vfit}) for the
lattice with $a_s=0.36$~fm.}
\label{fig:FWilScale}
\end{figure}

\begin{figure}
\vspace{0.25cm}
\centerline{\psfig{file=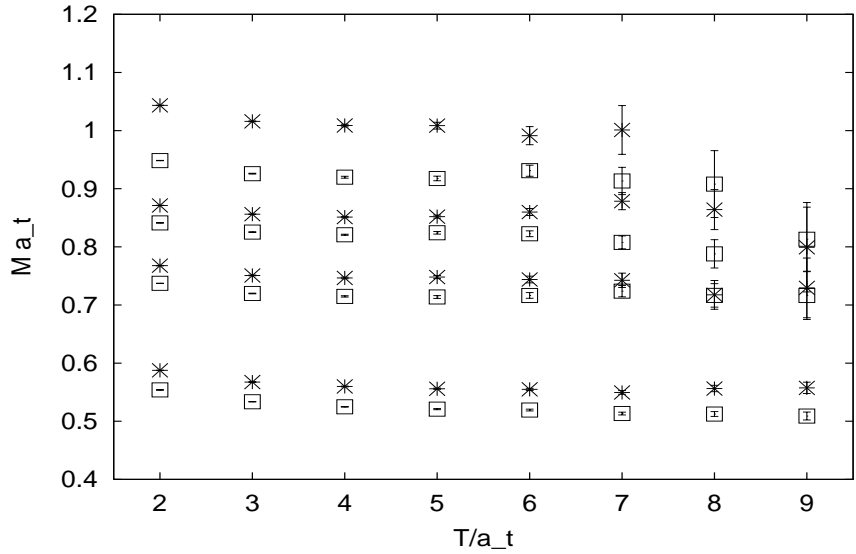,height=7.5cm,width=12.0cm}}
\vspace{0.50cm}
\caption{Effective mass plots for single electric ($*$) and
magnetic ($\Box$) gluelumps. There are four pairs of plots corresponding
to the four lattices, with the spatial spacing $a_s$ increasing from the
bottom of the figure to the top.}
\label{fig:GlueEmass}
\end{figure}

\begin{figure}
\vspace{0.25cm}
\centerline{\psfig{file=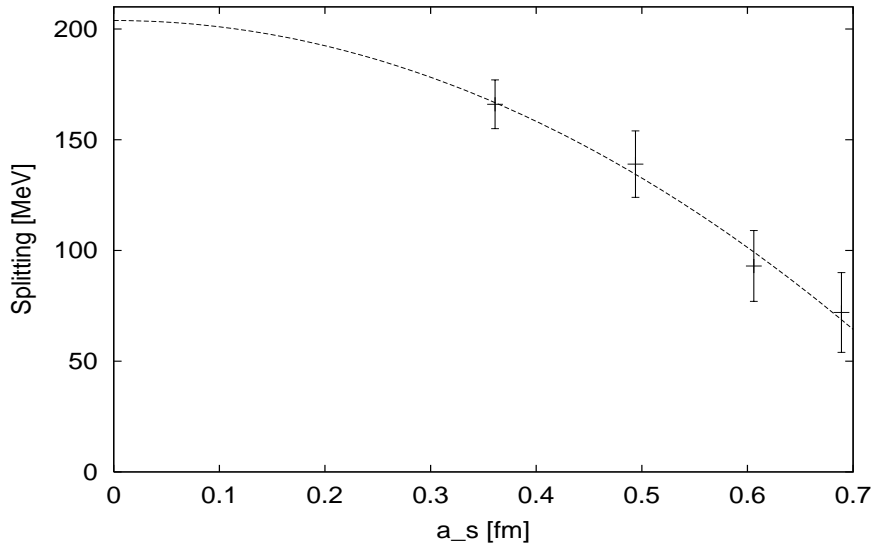,height=7.5cm,width=12.0cm}}
\vspace{0.50cm}
\caption{Gluelump physical mass splitting versus lattice spacing $a_s$.
The dashed curve shows a fit assuming $O(a^2)$ scaling violations.}
\label{fig:GlueSplitting}
\end{figure}

The spatial lattice spacing is then determined by fitting the
fundamental representation potential to the form
\begin{equation}
   V_{\rm fit}(R) = \sigma R - {b \over R}  + c ,
\label{Vfit}
\end{equation}
taking the physical value of the string tension to be
$\sqrt\sigma = 0.44$~GeV. The potentials were measured at on-axis
separations, as well as at off-axis separations $R/a_s = \sqrt2$,
$\sqrt3$, $\sqrt5$, $\sqrt8$, $\sqrt{13}$, $\sqrt{18}$, and $\sqrt{20}$.
Symmetric combinations of the shortest spatial paths connecting two
lattice points were used in the nonplanar Wilson loop calculations.
Results for the lattice with $a_s = 0.49$~fm are given in
Fig.~\ref{fig:FWilFit}, which shows fairly good rotational symmetry
restoration, thanks to tadpole improvement. A quantitative measure
of the symmetry breaking is obtained by comparing the simulation
results for the potential with the interpolation to the on-axis data
\begin{equation}
   \Delta V(R) \equiv
   {V_{\rm sim}(R) - V_{\rm fit}(R) \over \sigma R} .
\label{DeltaV}
\end{equation}
Results for $\Delta V$ at $R=\sqrt3 a_s$ for the four lattices
are shown in Table~\ref{table:params}. Unimproved actions
exhibit much larger rotational symmetry breaking effects
on such coarse lattices \cite{NormSU2}.

Representative plots of the time-dependent effective potential
\begin{equation}
   V(T,R) = -\ln\left( { W(T,R) \over W(T-1,R) } \right)
\end{equation}
are shown in Fig.~\ref{fig:FWilEmass}.
A reliable determination of the ground state potential in the
fundamental representation can be made, with excellent plateaus
in the effective mass plots, going out to propagation times near 1~fm,
even at separations as large as 2.5~fm.

The potentials from all four lattices
are plotted together in physical units in Fig.~\ref{fig:FWilScale}.
A Coulomb term is visible in this data, with fits to the on-axis
data yielding coefficients $b$ around 0.1 \cite{NormSU2}. Although the
fundamental potentials on these coarse lattices are dominated by the
confinement term, the results give some indication of fairly
good scaling behavior.

\subsection{Magnetic and electric gluelumps}

The effective mass plots for single electric and magnetic gluelumps
exhibit good plateaus, as shown in Fig.~\ref{fig:GlueEmass}.
The electric gluelump is known to have the larger mass \cite{MichaelAdj}.
The gluelump energy $M_{Qg}$ is not physical as it
must be additively renormalized due to the self energy of
the heavy quark.  However this renormalization should cancel
in the difference between the electric and magnetic gluelump energies.
A direct comparison of $2 M_{Qg}$ with the static potential
for a pair of adjoint quarks is also meaningful since the two quantities
have equal self energies.

Our results for the gluelump splittings on the four lattices are:
\begin{equation}
    M_{\rm elec} - M_{\rm mag} = \left\{
    \begin{array}{rc}
        166 \pm 11 \mbox{\ MeV,} \quad & a_s = 0.36 \mbox{\ fm,} \\
        139 \pm 15 \mbox{\ MeV,} \quad & a_s = 0.49 \mbox{\ fm,} \\
        93 \pm  16 \mbox{\ MeV,} \quad & a_s = 0.61 \mbox{\ fm,} \\
        72 \pm  18 \mbox{\ MeV,} \quad & a_s = 0.69 \mbox{\ fm,} \\
    \end{array}
    \right.
\label{gluesplit}
\end{equation}
and are plotted versus lattice spacing in Fig.~\ref{fig:GlueSplitting}.
For the sake of illustration, a fit assuming $O(a^2)$ scaling
violations yields a continuum estimate of
\begin{equation}
   M_{\rm elec} - M_{\rm mag} = ( 204 \pm 16 ) {\rm \ MeV},
   \quad \chi^2 / {\rm dof} = 0.29 ,
\end{equation}
while a fit assuming $O(a)$ scaling violations yields
\begin{equation}
    M_{\rm elec} - M_{\rm mag} = ( 273 \pm 29 ) {\rm \ MeV},
    \quad \chi^2 / {\rm dof} = 0.50 .
\end{equation}
These results are consistent with an estimate of the gluelump
splitting in SU(2) color by Jorysz and Michael \cite{MichaelAdj},
who found $M_{\rm elec} - M_{\rm mag} = 203 \pm 76$~MeV, using
a single lattice with a spacing of about 0.16~fm.

\subsection{Adjoint representation Wilson loops}

\begin{figure}
\vspace{0.25cm}
\centerline{\psfig{file=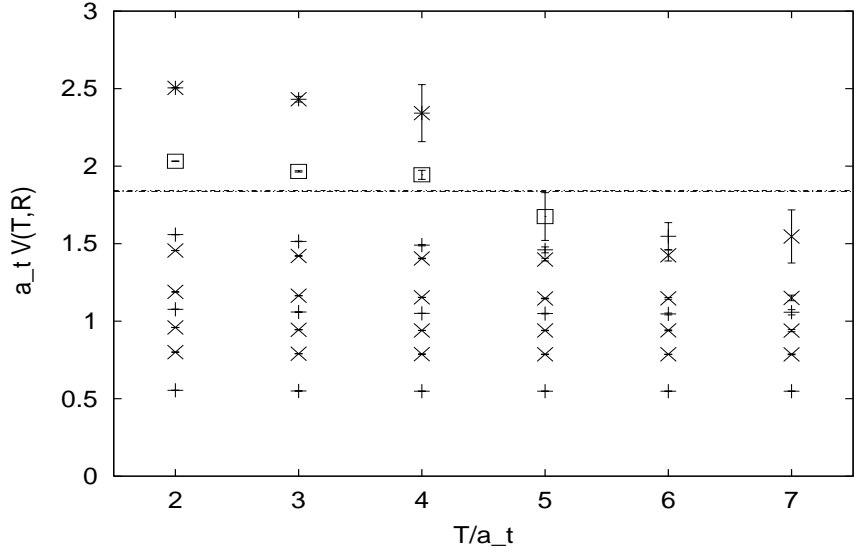,height=7.5cm,width=12.0cm}}
\vspace{0.50cm}
\caption{Adjoint Wilson loop effective mass plots for the lattice
with $a_s = 0.36$~fm and $a_t = 0.10$~fm. Plots are shown for
several values of the on-axis quark separation,
$R=1\mbox{--}3$ ($+$), $R=4$ ($\Box$) and $R=5$ ($*$), as well
as at some off-axis points ($\times$).
The dashed lines show the $1\sigma$ limits for twice the mass
of the magnetic gluelump $2 M_{Qg}$ on this lattice.}
\label{fig:AWilEmass36}
\end{figure}

\begin{figure}
\vspace{0.25cm}
\centerline{\psfig{file=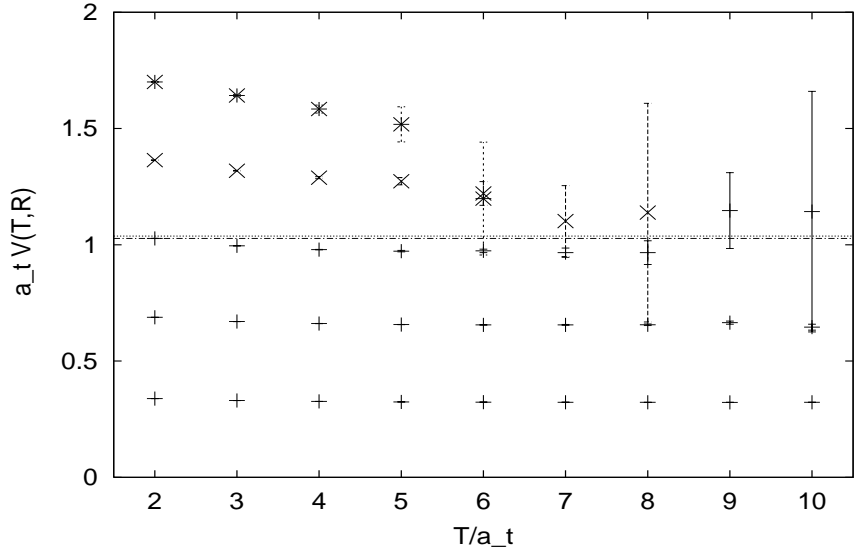,height=7.5cm,width=12.0cm}}
\vspace{0.50cm}
\caption{Adjoint Wilson loop effective mass plots for the lattice
with $a_s = 0.49$~fm and $a_t = 0.072$~fm. Plots are shown for
$R=1\mbox{--}3$ ($+$), $R=4$ ($\times$), $R=5$ ($*$).}
\label{fig:AWilEmass49}
\end{figure}

The determination of the ground state potential in the adjoint
representation is much more difficult than in the fundamental case, due
at least in part to a much larger overall energy scale in the
adjoint channel. Effective mass plots for the adjoint potential on two
lattices are shown in Figs.~\ref{fig:AWilEmass36} and \ref{fig:AWilEmass49}.
Notice that the temporal spacing is smaller in Fig.~\ref{fig:AWilEmass49},
allowing one to more clearly see that plateaus in the effective masses at
large separations have not been reached. The trend in the effective mass
plots at large $R$ is not inconsistent with the suggestion \cite{HDT}
that string breaking occurs at propagation times of about 1~fm.

\begin{figure}
\vspace{0.25cm}
\centerline{\psfig{file=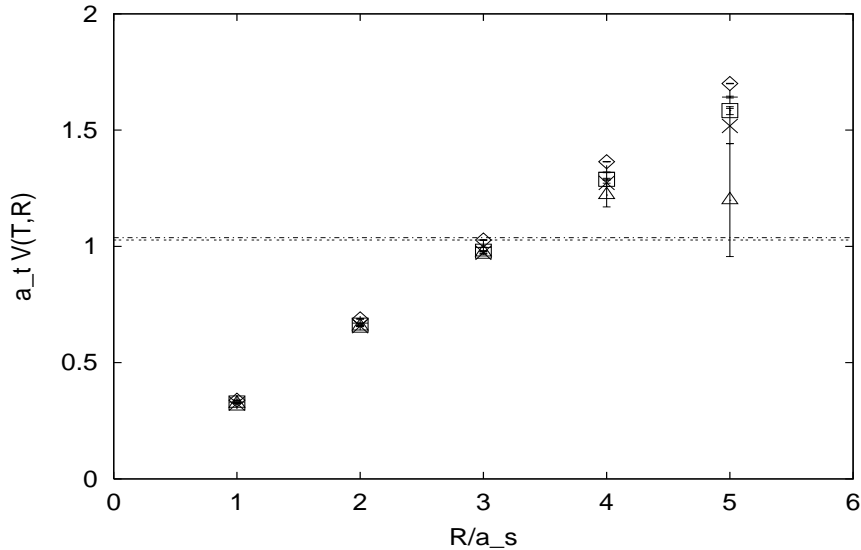,height=7.5cm,width=12.0cm}}
\vspace{0.50cm}
\caption{Adjoint Wilson loop static potential vs $R$ at various fixed
propagation times, for the lattice with $a_s=0.49$~fm and $a_t=0.07$~fm:
$T/a_t=2$ ($\diamond$), 3 ($+$), 4 ($\Box$), 5 ($\times$), 6 ($\triangle$).
The dotted lines show $1\sigma$ limits for $2 M_{Qg}$ (magnetic).}
\label{fig:AWilVRTfixed}
\end{figure}

A typical procedure followed in the literature on string breaking is to
estimate the ground state potential $V(R)$ as equal to the effective
potential $V(T,R)$ at a small value of $T$, especially at large $R$,
given the poor signal-to-noise in this region. The effect of choosing
different fixed values of $T$ for the determination of the potential can
be seen by plotting $V(T,R)$ versus $R$, for several choices of $T$.
We show our data in this way for one lattice in Fig.~\ref{fig:AWilVRTfixed}.
There is a clear trend for the ``potential'' to flatten as $T$ is increased,
and this trend continues until the signal at large $R$ is lost in the noise.
The limitation to such small propagation times $T \alt 0.4$~fm at large
separations introduces a significant systematic error in assessing whether
the potential saturates.

\begin{figure}
\vspace{0.25cm}
\centerline{\psfig{file=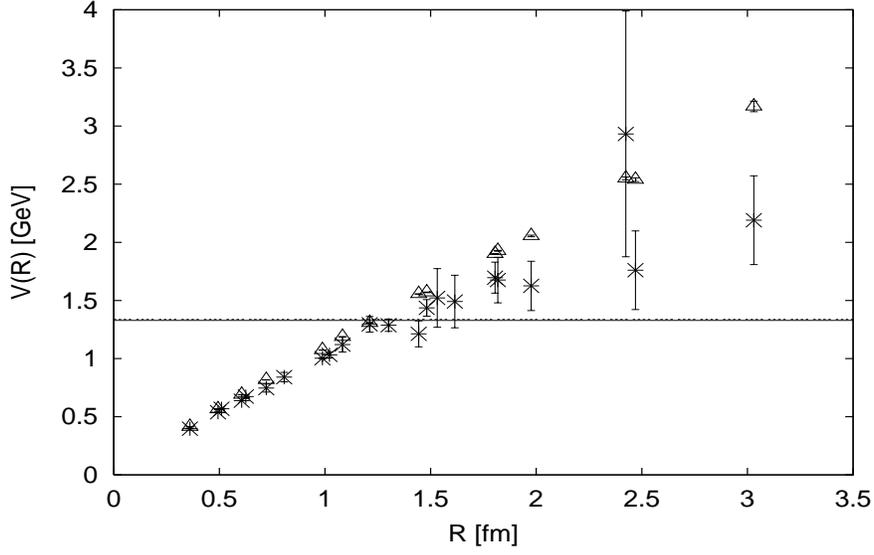,height=7.5cm,width=12.0cm}}
\vspace{0.50cm}
\caption{Comparison of the adjoint ($*$) and fundamental ($\triangle$)
potentials in physical units from all four lattices. The adjoint potential
has been multiplied by a factor of $3/8$ and shifted vertically to agree
with the fundamental potential at $R \approx 0.5$~fm. The dashed lines show
$1\sigma$ limits for $2 M_{Qg}$ (magnetic), after being rescaled and
shifted vertically by the same amount as the adjoint potential (here using
the results only at $a_s=0.36$~fm).}
\label{fig:Casimir}
\end{figure}

It is also useful to compare the fundamental potential with the
adjoint one. In Fig.~\ref{fig:Casimir} we plot the two potentials
in physical units from all four lattices, where we rescale the
adjoint potential by $3/8$, the ratio of SU(2) Casimirs for the
two representations. There is good evidence for screening
of the adjoint potential, compared to simple models of Casimir
scaling \cite{MichaelAdj}, but again the extent of the screening
(or possible saturation) cannot be reliably determined, due to the
limitation to short propagation times.

\subsection{Gauge-fixed quark-antiquark correlator}

\begin{figure}
\vspace{0.25cm}
\centerline{\psfig{file=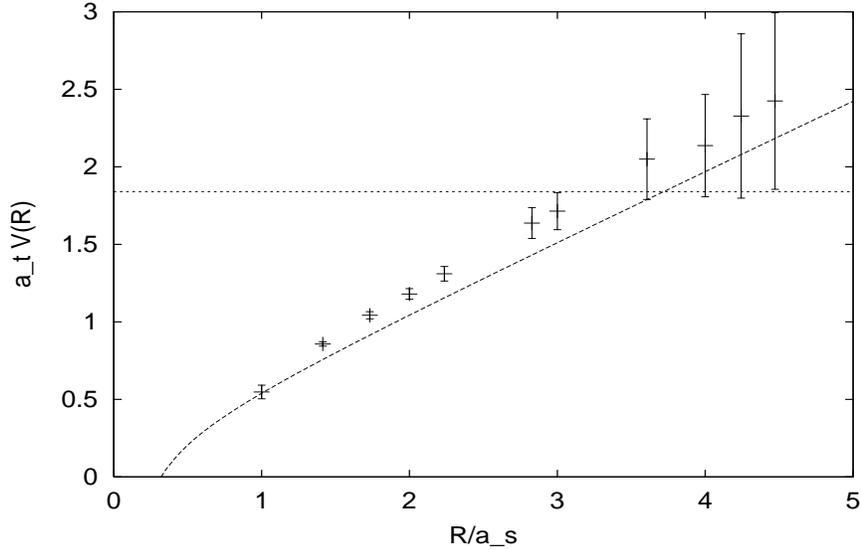,height=7.5cm,width=12.0cm}}
\vspace{0.50cm}
\caption{Adjoint potential computed from gauge-fixed static quark
propagators on the lattice with $a_s=0.36$~fm and $a_t=0.10$~fm.
The dotted line shows $2 M_{Qg}$ (magnetic),
and the dashed curve shows the result of a fit to the potential
computed from the adjoint Wilson loop on the same lattice
and at comparable propagation times.}
\label{fig:GaugedStatic}
\end{figure}

The correlation function  between a pair of static adjoint quark
propagators (cf. Eq. (\ref{GPoly})) was calculated in Coulomb gauge, in order
to study a state without an explicit string of links connecting the heavy
quarks. Similar correlators were suggested for observing string breaking in
Refs.~\cite{Burger,DeTar}. Results for the effective potential
defined from $G_{\rm Poly}(T,R)$ are shown for one lattice
in Fig.~\ref{fig:GaugedStatic}. The results obtained from this
correlator agree well with the Wilson loop estimate of the potential,
obtained at similar propagation times, giving neither a better
nor a worse indication of string breaking.

\subsection{Gluelump-gluelump correlators}

\begin{figure}
\vspace{0.25cm}
\centerline{\psfig{file=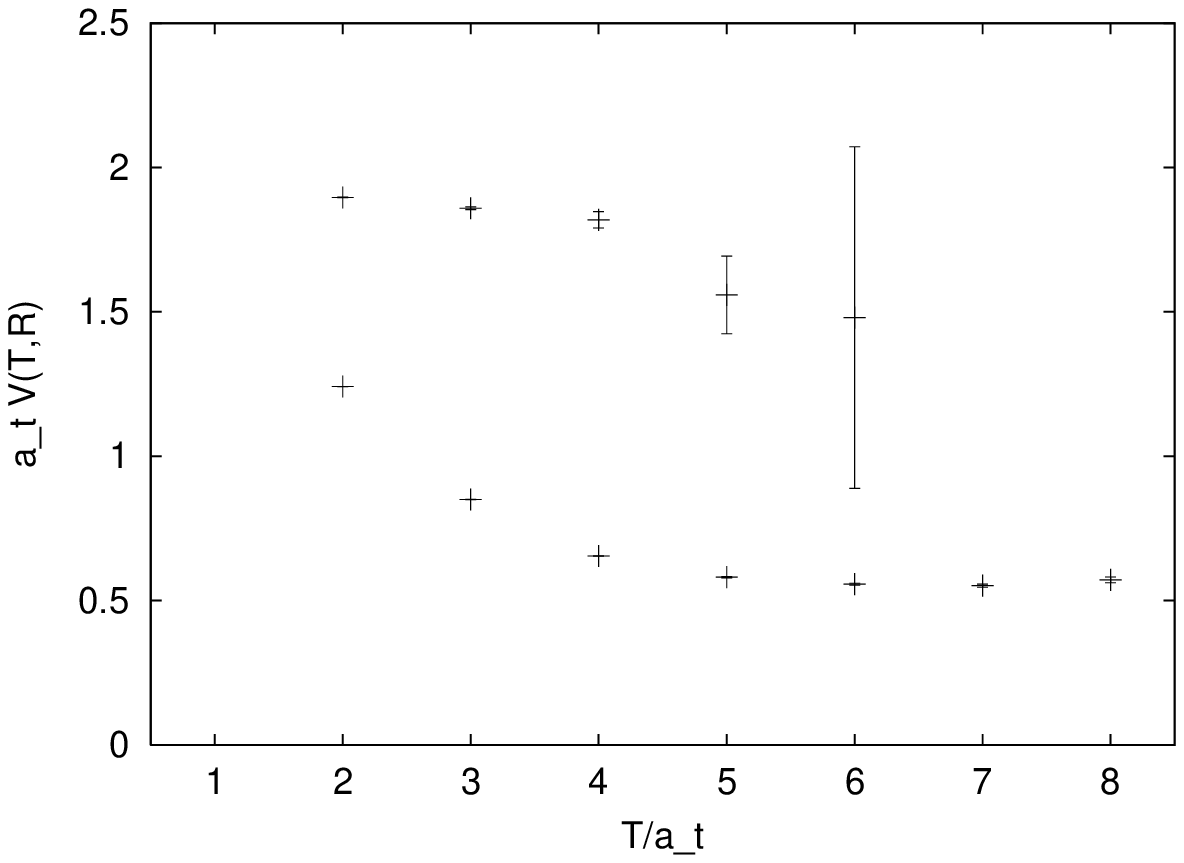,height=7.5cm,width=12.0cm}}
\vspace{0.50cm}
\caption{Magnetic gluelump-gluelump effective masses for the
lattice with $a_s=0.36$~fm and $a_t=0.10$~fm, for two separations:
$R=1$ (lower points) and $R=4$ (upper points).}
\label{fig:GGEmass}
\end{figure}

\begin{figure}
\vspace{0.25cm}
\centerline{\psfig{file=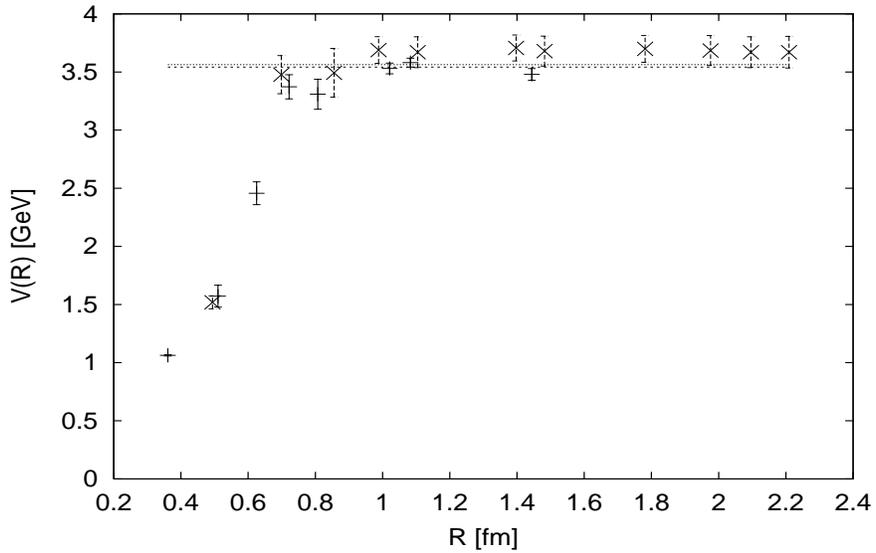,height=7.5cm,width=12.0cm}}
\vspace{0.50cm}
\caption{Static potential estimates from the magnetic gluelump-pair
correlator in physical units (with additive energy shifts):
$a_s=0.36$~fm ($+$) and $a_s=0.49$~fm ($\times$).}
\label{fig:GGScale}
\end{figure}

Representative effective mass plots for the magnetic gluelump-gluelump
correlator (cf. Eq.~(\ref{GGPair})) for one lattice are shown in
Fig.~\ref{fig:GGEmass}. At smaller separations the signal is clear but
contains large excited state contributions; at larger separations the
signal degrades, but the data at small $T$ show more of a plateau.
The resulting potentials from two lattices are compared in physical
units in Fig.~\ref{fig:GGScale}.

We also used a standard variational method \cite{MichaelAdj} to estimate
the state of lowest energy in the $2\times2$ basis of states composed of
a pair of heavy adjoint quarks connected to each other by a string of
links (adjoint Wilson loop $W_{\rm adj}$), and a pair of gluelumps.
Consider the corresponding $2\times2$ correlation matrix $C_{ij}$
(cf. Eqs. (\ref{GGPair})--(\ref{GWG}))
\begin{equation}
   C_{ij}(T,R) = \left(
   \begin{array}{cc}
     \left\langle W_{\rm adj}(T,R) \right\rangle
   & \left\langle G_{GW}(T,R) \right\rangle \\
     \left\langle G_{WG}(T,R) \right\rangle
   & \left\langle G_{GG}(T,R) \right\rangle \\
   \end{array}
   \right) .
\end{equation}
Representing the two basis states by $\vert \phi_i(R) \rangle$, the
correlation matrix $C_{ij}(T,R)$ is written as a transfer matrix
\begin{equation}
   C_{ij}(T,R) = \langle \phi_i(R) \vert e^{-HT} \vert \phi_j(R) \rangle .
\end{equation}

We first find a linear combination $\vert \Phi(R) \rangle$ of basis states
\begin{equation}
   \vert \Phi(R) \rangle = \sum_i a_i(R) \vert \phi_i(R) \rangle
\end{equation}
which maximizes
\begin{equation}
   \lambda(T^*,R) = { \langle \Phi(R) \vert e^{-HT^*} \vert \Phi(R) \rangle
   \over \langle \Phi(R) \vert \Phi(R) \rangle } .
\label{lambdaT}
\end{equation}
This requires the solution of the eigenvalue problem
\begin{equation}
   C_{ij}(T^*,R) a_j(R) - \lambda(T^*,R) C_{ij}(0,R) a_j(R) = 0 .
\label{VarEigen}
\end{equation}
We choose to optimize the variational state by solving Eq. (\ref{VarEigen})
at a small time $T^*$, otherwise numerical instabilities may arise
due to large statistical errors in $C_{ij}(T^*,R)$, especially
at large $R$.

We then evolve the correlation function Eq. (\ref{lambdaT}) for the
variational state to a larger time $T$, in order to filter out our final
estimate of the ground state energy $\lambda(T,R) = e^{-E_0(R)T}$.
The overlaps
$c_i^2(R) = \langle \phi_i(R) \vert {\cal O}(R) \rangle^2
/ \langle \phi_i(R) \vert\phi_i(R) \rangle$ of the
basis states $\vert \phi_i(R) \rangle$ on the ground state
$\vert {\cal O}(R)\rangle$ can be estimated according to
\begin{equation}
   c^2_i(R) = { C_{ii}(T,R) \over \lambda(T,R) C_{ii}(0,R) } ,
\label{GroundOverlap}
\end{equation}
at sufficiently large $T$ (Eq. (\ref{GroundOverlap})
at finite $T$ provides an upper bound on the overlaps).

\begin{figure}
\vspace{0.25cm}
\centerline{\psfig{file=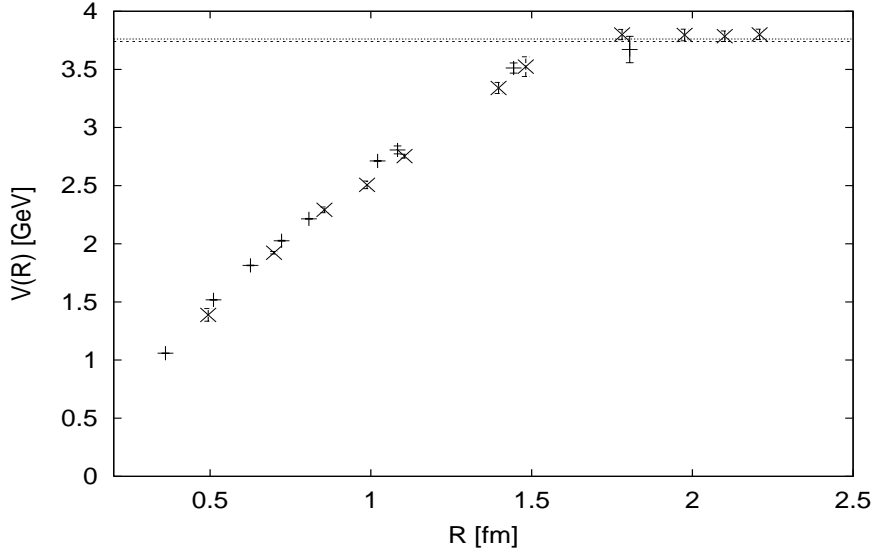,height=7.5cm,width=12.0cm}}
\vspace{0.50cm}
\caption{Variational estimate of the ground state energy in physical
units for two lattices (after additive shifts in the energies):
$a_s=0.36$~fm ($+$) and $a_s=0.49$~fm ($\times$). The trial
state was typically determined at $T^*=a_t$, which was then propagated
to a time $T \approx 4 a_t$ to obtain the results shown in the
figure. Also shown are the $1\sigma$ lines for $2M_{Qg}$ (magnetic).}
\label{fig:DiagScale}
\end{figure}

\begin{figure}
\vspace{0.25cm}
\centerline{\psfig{file=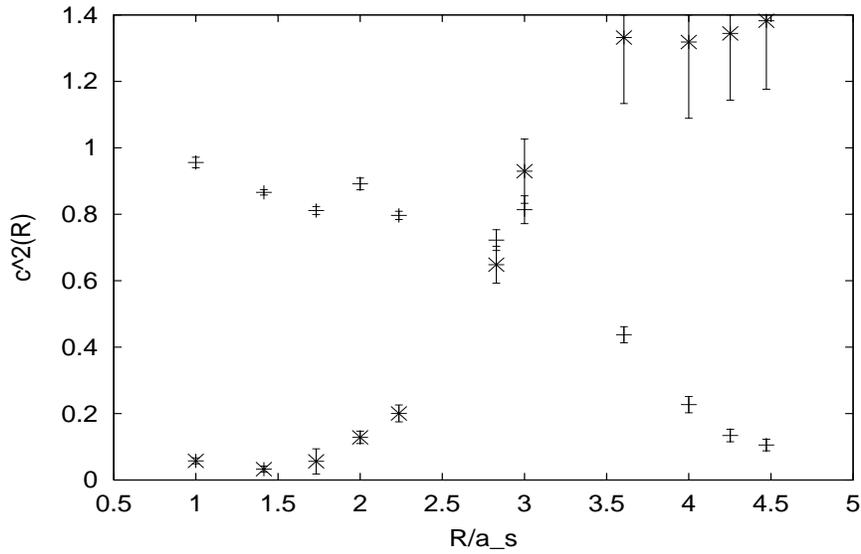,height=7.5cm,width=12.0cm}}
\vspace{0.50cm}
\caption{Diagonalization results for the lattice with $a_s=0.49$~fm,
showing the overlaps of the two basis states with the
ground state as functions of $R$, estimated from Eq. (\ref{GroundOverlap}):
Wilson loop ($+$) and gluelump-pair ($*$).}
\label{fig:DiagOverlaps}
\end{figure}

The results of this diagonalization procedure are as might be
expected. Figure~\ref{fig:DiagScale} shows the estimate of the
ground state potential in physical units from two lattices.
The estimated overlaps of the Wilson loop and gluelump-pair states
with the variational estimate of the ground state
are shown in Fig.~\ref{fig:DiagOverlaps}. There is a rapid
cross-over in the ground state as determined in this basis, from
the Wilson loop at smaller $R$ to the gluelump-pair state at larger $R$.
The rapid cross-over suggests that the gluelump-pair state becomes
dominated by disconnected contributions when the energies of the
states near $2M_{Qg}$.

\section{Summary and Further Discussion}
In this paper we made a through analysis of adjoint quark physics in
the context of string breaking. Three trial states were
investigated as candidates for observing string breaking:
the adjoint Wilson loop, a pair of gluelump propagators, and a pair of
gauge-fixed static quark propagators. The fundamental representation
potentials were used to measure the lattice parameters, and electric
and magnetic gluelump masses were calculated in order to set the
energy scale for string breaking. A number of techniques were
used to maximize the efficiency of the calculations, including fuzzing,
variance reduction, and fast Fourier accelerated gauge fixing.
Particularly important was the use of coarse, highly anisotropic
lattices from tadpole-improved actions. The lattice with $a_s =0.36$~fm and
$a_t=0.10$~fm for example gives an improvement in computational efficiency
of some two orders of magnitude compared to simulations of
adjoint quarks done in Ref.~\cite{PhilAdjoint4D} on lattices
with spacings of about 0.1~fm.
Large lattice anisotropies provided more data points for analysis of the
Euclidean time evolution of correlation functions. This proved to be
especially important in analyzing adjoint Wilson loops for even moderate
physical values of $R$, due to the rapid decay of the signal.

Diagonalization in a two state basis, corresponding to adjoint
Wilson loops and the gluelump-pair operators, reveals a static
potential which saturates at $2 M_{Qg}$ near $1.5$ fm.
Similar string breaking distances have been suggested for full
QCD \cite{Sommer}, and have been observed in other theories, including
three-dimensional QCD with dynamical fermions \cite{HDT}. At small quark
separations the potential rises linearly, with a slope of about
$\frac{8}{3}$ of the fundamental potential, consistent with Casimir scaling.
Saturation in the potential obtained from the diagonalization occurs over
a very small range of separations $R$. These results are in qualitative
agreement with recent calculations done on fine lattices using unimproved
actions \cite{Stephenson,PhilAdjoint3D,PhilAdjoint4D}.
In addition, results obtained here using correlations between gauge-fixed
static quark propagators agree well with the Wilson loop estimate of the
potential, giving neither a better nor a worse indication of string breaking.

As discussed in Sect.~I these results, taken at face value, might be
interpreted as evidence that Wilson loops are not suitable for studies of
string breaking, and that string breaking can be readily observed using
operators that explicitly contain fields for the light particles. However we
raised several cautionary observations about this viewpoint. In particular,
we pointed out in Sect.~I that operators which explicitly generate a pair of
light valence particles {\it automatically\/} exhibit potentials
that saturate. In the case of QCD with fundamental representation quarks,
this means that operators of this type would demonstrate ``string breaking''
even in the quenched theory. This is particularly clear when the
correlation function receives disconnected contributions. In the case
of the gluelump-pair correlator (cf. Eq. (\ref{GGPair})), we have
\begin{equation}
   \langle 0 \vert G^\dagger(T;R) G(T;0) \vert 0 \rangle =
   \langle 0 \vert G(T;0) \vert 0 \rangle^2  + \ldots ,
\label{VacSat}
\end{equation}
where $\vert 0 \rangle$ is the vacuum state, and where the ellipsis denotes
the contributions of non-vacuum insertions between the two operators in
the correlator. Hence one is guaranteed to find an effective mass of
$2 M_{Qg}$ at modest $R$ using this correlation function. This also helps
to explain the rapid cross-over that is observed in variational
calculations using Wilson loops and such states
\cite{PhilScalar,Knechtli,Stephenson,PhilAdjoint3D,PhilAdjoint4D},
when the energies of the states approach the broken string energy.

Exactly the same behavior must occur if operators that explicitly generate
light valence quarks are used in simulations of QCD with fundamental
representation quarks. This can also be seen from some recent work using
such operators in quenched QCD \cite{Koniuk}.
Thus operators of the type that have recently been proposed in
Refs. \cite{PhilScalar,Knechtli} for observing string breaking
will yield qualitatively the same behavior in both quenched and
unquenched QCD.

This raises the important question of exactly how one defines the goal
of ``observing string breaking.'' Perhaps one can advocate two points of view.
From one point of view, one would say that correlation functions give spectral
information only: the ground state energy versus quark separation. In this
view such information is useful, for example, in characterizing the
effects of quenching on meson-meson interaction energies
\cite{KoniukMeson,Fiebig}, but not in relation to processes such as
hadronization, which can only occur in the presence of dynamical
fermions. Then one might say that the difficulty in seeing ``string breaking''
with the Wilson loop at very large separations indicates nothing more than that
this operator has a poor overlap with the ground state in this regime.
If spectral information is all that one is interested in, then one might
not be too concerned that a particular operator shows qualitatively similar
behaviour in the quenched and unquenched theories.

A second point of view, which we raised in Sect.~I, is that one can use
certain correlation functions to make an analogy with hadronization. One
might argue that this may be done by considering operators that do not
generate light valence particles in the trial state. In particular,
the operator should exhibit ``string breaking'' only in unquenched QCD
(here considering the theory with fundamental representation quarks).
The Wilson loop is such an operator, while operators which explicitly
generate light valence particles are not. Moreover,
if a goal of string breaking studies is to observe a distinctive feature
of the effects of sea quarks, then one should only consider
observables that distinguish between the quenched and unquenched theories.
In a sense the first point of view discussed above is concerned with
``static'' spectral information, that is qualitatively similar in the
quenched and unquenched theories, while the second point of view makes
contact with a ``dynamical'' process that is unique to the unquenched theory.

In making contact with hadronization one is interested in using the Wilson
loop to measure the potential for separations $R$ not much larger than the
distance at which the string breaks, $R \approx 1.5$~fm, since in the actual
physical process the original quarks never get to larger separations with the
string intact. In this region the overlap of the Wilson loop with the ground
state appears to be appreciable, judging from Fig.~\ref{fig:DiagOverlaps},
and from results presented in Ref. \cite{HDT}.
However one must still push the calculation to propagation times of about
1~fm, in order to properly resolve the broken string. This is where the
difficulty actually lies in observing string breaking using Wilson loops.

To date most simulations of full QCD have been done on lattices with
relatively fine spacings, making it computationally very challenging
to reach the length scales $R\approx1.5$~fm and propagation times
$T\approx1$~fm relevant to string breaking.
The use of coarse lattices with improved actions
allows a much more efficient probe of this regime, as was recently
demonstrated by one of us in Ref. \cite{HDT}, where this approach
was used to resolve string breaking with Wilson loops in unquenched QCD
in three dimensions. An increase in computational efficiency of some two
orders of magnitude is possible using lattices with spacings between 0.3~fm
and 0.4~fm, compared to most simulations that have been done so far in
unquenched QCD. Some work in this direction has recently been reported
in Ref. \cite{Eichten}.

Unfortunately calculations of the adjoint Wilson loop in this string breaking
regime did not prove to be feasible in this paper, with propagation times
limited to well below 1~fm, even with coarse lattices. On the other
hand it is clear that adjoint Wilson loops exhibit a potential that
progressively ``flattens'' at longer propagation times. Moreover,
the trend in the effective mass plots from the adjoint Wilson loop at large
$R$ is not inconsistent with the suggestion that string breaking would be
observed at propagation times of about 1~fm.

In this context it is interesting to estimate the size of the adjoint
Wilson loop signal relative to the fundamental one, and to compare the
computational cost of these simulations to those of unquenched QCD. If we
assume roughly Casimir scaling of the potential just below the string
breaking distance, then the ratio $W_{\rm adj}/W_{\rm fund}$ of the adjoint
to the fundamental Wilson loops in SU(3) color goes like
\begin{equation}
   W_{\rm adj} / W_{\rm fund}
   \approx \exp\left[ - (\case{9}{4} - 1) \sigma RT \right] \approx 10^{-4} ,
\end{equation}
using $\sqrt\sigma=0.44$~GeV for the fundamental representation
string tension, and $R\approx1.5$~fm and $T\approx1$~fm
for the scales relevant to string breaking (the ratio is yet smaller,
by an order of magnitude, in SU(2) color). This is to be compared
with the roughly two orders of magnitude increase in the cost of
simulating dynamical quarks compared to quenched simulations.
(Note that the gluelump-gluelump correlator
\cite{Stephenson,PhilAdjoint3D,PhilAdjoint4D} does not show
a comparable suppression of the signal, which is entirely a consequence
of not removing the disconnected contributions from the correlator).
Hence, while the adjoint representation is interesting as a probe of
confinement and supersymmetric physics, it is not a
cost effective means of mimicking hadronization in full QCD.

On the other hand the results presented here do lend support to the general
picture that Wilson loops should in fact exhibit ``string breaking,'' as an
analogy to hadronization. Moreover string breaking should be accessible in
real QCD with the computational power currently available in large scale
simulation environments, especially if coarse lattices with improved
actions are used.

\acknowledgments
We thank E. Eichten, B. Jennings, J. Juge, G. Moore and R. Woloshyn for
fruitful conversations. This work was supported in part by the Natural
Sciences and Engineering Research Council of Canada.


\end{document}